\documentclass[prl,a4paper,twocolumn,showpacs]{revtex4}
\usepackage{graphicx}
\newcommand{\be}{\begin{equation}}
\newcommand{\ee}{\end{equation}}
\newcommand{\bea}{\begin{eqnarray}}
\newcommand{\eea}{\end{eqnarray}}
\newcommand{\ba}{\begin{array}}
\newcommand{\ea}{\end{array}}
\newcommand{\p}{\partial}
\newcommand{\al}{\alpha}
\newcommand{\ra}{\rangle}
\newcommand{\la}{\langle}

\begin{document}
\title{The Behavior of Electronic Interferometers in the Non-Linear Regime}
\author{I. Neder and  E. Ginossar}
\affiliation{Department of Condensed Matter Physics, The Weizmann
Institute of Science, Rehovot 76100, Israel}
\email{izhar.neder@weizmann.ac.il}
\date{\today}
\begin{abstract}
We investigate theoretically the behavior of the current
oscillations in an electronic Mach-Zehnder interferometer (MZI) as
a function of its source bias. Recently, The MZI interference
visibility showed an unexplained lobe pattern behavior with a
peculiar phase rigidity. Moreover, the effect did not depend on
the MZI paths difference. We argue that these effects may be a new many-body manifestation of
 particle-wave duality of quantum mechanics. When biasing the interferometer sources beyond the linear response regime, quantum shot-noise (a particle phenomena) must affect the interference pattern of the electrons that creates it, as a result from a simple invariance argument. An approximate solution of the interacting Hamiltonian indeed shows that the interference visibility has a lobe pattern with applied bias with a period proportional to the average path length and independent of the paths difference, together with a phase rigidity.

\end{abstract}
\maketitle
In the last
two decades, electron interferometers became a primary tool in
mesoscopic physics, for investigating quantum coherence of
 transport in semi-conductors, and to measure and control novel
quantum effects \cite{Tim87,Yac96,Buk98,Jiy03,Sam03,Ned07c}. Their behavior is
understood so far only in the linear response regime, where a
small bias is put on their sources, and the non-interacting
picture is valid. In the non-linear regime several electrons are present inside the interferometer at a given time and may form, in contrast to the optical interferometers, non-trivial many-body correlations due to Coulomb interaction. Indeed, new experiments in the non-linear regime demonstrated that the Landauer-Buttiker formalism \cite{But88} seems to break down, and the interference pattern showed new peculiar behavior. Recently, an unexpected interference behavior of a Mach-Zehnder
interferometer (MZI) was reported \cite{Ned06, Rou07}, in which the visibility (proportional to the observed amplitude of the Aharonov-Bohm (AB) oscillations of the drains current) evolved in a lobe pattern with increasing the source bias, with zero visibility between the lobes, and a phase independence on the bias inside each lobe ("phase rigidity"). Similar visibility lobes were observed earlier in a two-terminal closed interferometers
\cite{Yac96,Wie03}. In the MZI case, the lobe pattern could not be explained using any non-interacting picture, mainly because the lobes were very robust against induced asymmetry between the two path, and did not depend at all on their length difference. Recent explanations proposed for this effect using
Bosonization techniques \cite{Suk07,Cha07,Lev08} imposed interactions
between electrons that were geometry dependent, somewhat in
contrast to the robustness of the experimental observation.
Moreover, these propositions do not explain the total phase
rigidity seen in the experiment \cite{Ned06}.

Here we show that these lobes and phase rigidity are indeed very
reasonably a non-perturbative result of the Coulomb interaction. We argue that in the non-linear regime, this interaction causes the extra electrons inside a two path interferometer to form a correlated state which is rooted in their quantum behavior; as electrons behaves both as particles and as waves, the phase of an interfering electron gets discrete ``quantum kicks" according to the possible occupations of the additional electron states in the two arms. In some distinct source voltages those phase fluctuations cause the vanishing of the interference, which reappear at lower or higher voltages, leading to a lobe pattern in the visibility.
\begin{figure} [h] \centering
\includegraphics[scale=0.25]{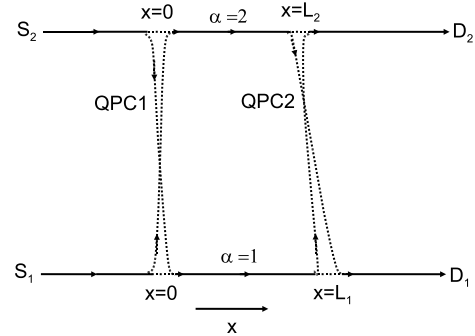}
\caption{Schematic of the MZI geometry, with generally asymmetric
two paths.} \label{fig1}
\end{figure}
The basis for our derivation is a general principle for mesoscopic devices, which we
term Buttiker-Levinson argument (BLA): changing the chemical
potential of all the sources by the same amount, should not change
the transmission from any source to any drain. It is generally relevant for any energy-dependant mesoscopic devices \cite{Chr96},and in particular for the MZI, being two-path interferometer formed by one dimensional edge channels in the quantum Hall effect regime, since the interaction between the electrons inside the paths is unscreened by the gates nearby. Surprisingly, when applied to electron interferometers in the non-linear regime, this
simple argument is violated by non-interacting models \cite{Ned06}.

A seemingly natural solution to this inconsistency is to correct the non-interacting models by taking into account a mean field approximation (MFA)\cite{Chr96,Wie03,Ned06} in which the bias induces a
static mean charging in each of the interferometer arms which causes a phase shift due to Coulomb interaction. Experimentally however, this approximation failed for the MZI as the lobes were independent on the two arm length difference \cite{Ned06}. We now further argue that this MFA correction is never valid in the MZI, since whenever it is significant, quantum fluctuations are strong and cannot be neglected. As the phase shift must restore the BLA validity, the phase shift correction in the MZI must be given  by $e^{i2\pi\la N_1-N_2\ra}$ \cite{Fri52}, where $N_\alpha$ is the excess number of electrons in path $\alpha$ due
to the source bias. Due to the applied bias this number of
electrons and the phase in each arm fluctuate \emph{binomially}
(quantum shot-noise) with a large variance and, most importantly,
with strong negative correlation between the two arms. Therefore
the induced phase shift cannot be approximated simply by the
average charging since $\la e^{i\delta\phi(N)}\ra\neq e^{i\la
\delta\phi(N)\ra}$, as was suggested in \cite{Wie03}. Rather, in this situation of few
electrons inside the MZI generally $[N,\delta\phi]\neq0$, and non-linear interactions may cause new many-body coherent states analogous to squeezed spin states \cite{Kit91}. Our situation is shown below to be closely related
to another non-linear system, a biased finite 1D detector channel coupled a MZI arm \cite{Ned07b,Ned07a}.

Our starting point is the non-interacting Hamiltonian, written for
spinless or spin polarized electrons in the MZI drawn in
Fig. \ref{fig1}. It is expressed by the creation and annihilation operators
$c_k^{(S_\al)},c_k^{(S_\al)\dag}$ for the {\em incoming} extended
single-particle energy states from the two sources $S_{1,2}$, 
\be \label{eq:Ham0}
H_0=\sum_{\alpha=1,2}{\sum_{k}{(\epsilon(k)-\mu_{S_\alpha})c^{(S_\al)\dag}_kc_k^{(S_\al)}}}.
\ee In this model a linear relation can be established between the
{\em incoming} and {\em outgoing} $k$-operators from the sources
and to the drains, respectively \be
c_k^{(D_\al)}=\sum_{\beta=1,2}{s^{(MZI)}_{k,\alpha\beta}c_k^{(S_\beta)}}.
\ee where the s-matrix for the particular MZI geometry can be
calculated from the product $s^{(MZI)}=s^{(QPC2)}\cdot
s^{(\delta\varphi)}\cdot s^{(QPC1)}$ with the three s-matrices
defined as ($a=1,2$) \bea \label{QPCALPHA} && s^{(QPC(a))}=
\left(\begin{array}{cc}
ir_a & t_a \\
t_a &  ir_a
\end{array}\right), \\ \nonumber
&& s^{(\delta\varphi)} =
\left(\begin{array}{cc}
e^{ikL_1} & 0 \\
0 &  e^{ikL_2}
\end{array}\right)
\eea where $r_a$ and $t_a$ the reflection and transmission
amplitude of QPC(a), and $L_{1,2}$ are the two MZI path lengths.
We can define the annihilation operator $c_k^{(\alpha)}$ of an
electron moving in arm $\al$ of the MZI as obeying the relations
\bea \label{cks}
c_k^{(\alpha)}&=&\sum_{\beta=1,2}{s^{(QPC1)}_{\alpha\beta}c_k^{(S_\beta)}}\\ \nonumber
 c_k^{(D_\alpha)}&=&\sum_{\beta=1,2}{
\left(s^{(QPC2)}\cdot s^{(\delta\varphi)}\right)_{\alpha\beta}
c_k^{(\beta)}}. \eea Considering the local nature of
the interactions and of the device, we choose to express the current
in the drains using the local operators after QPC2,
$\psi_{(D_\alpha)}(x,t)=\sum_{k}e^{ikx}c_k^{(D_\alpha)}(t)$. The
resulting chemical potential at any drain $D_\al$, $\mu_{D_\alpha}$, is proportional to the
average current to the drain,
$\mu_{D_\alpha}=\frac{h}{e}\langle I_{D_\alpha}\rangle$, and since our problem is
stationary in time it can be written (linearizing $\epsilon(k)$) using this operator $\psi_{(D_\al)}(x,t)$
at $t=0$ \be \label{eq:mud2}
\mu_{D_\al}=\frac{h}{e}\la
I_{D_\al}\ra=hv_g \left\langle
\psi^{\dag}_{(D_\al)}(L_\al,0)\psi_{(D_\al)}(L_\al,0)\right\rangle
\nonumber
\ee where $v_g$ is the group velocity at the Fermi energy.
In the non-interacting model Eq. \ref{eq:mud2} can be
written as
$\mu_{D_\al}=\mu_{0,\al}+\Re(\mu_{\varphi,\al}e^{i\varphi_0})$,
with $\mu_{0,\al}$ a real parameter and $\mu_{\varphi,\al}$ a
complex pre-factor of the phase ($\varphi_0=k_F\Delta L$)
dependant term. The second term is
 proportional to $\int_{\mu_{S_1}}^{\mu_{S_2}}{\cos(\varphi_0+\frac{\Delta L}{\hbar
v_g}(\epsilon-E_F)) d\epsilon}=\left[2\hbar v_g\sin(\frac{\Delta
L\Delta\mu}{\hbar v_g})\right]\cdot\cos(\varphi_0+\frac{\Delta
L}{\hbar v_g}(\bar{\mu}-E_F))$, with $E_F$ the Fermi energy at
zero bias. It violates BLA through the explicit dependance on the average source bias $\bar\mu-E_F$.

Since, as stated above, a MFA correction is inadequate due to the
quantum fluctuations, we are forced to return to the Hamiltonian,
and introduce additional non-linear Coulomb terms (whose range we assume here to be longer then the MZI arms)
to take care of the BLA
\be \label{eq:NewHam}
H=H_0+\sum_{\alpha=1,2}\frac{e^2}{2C_\alpha}N^2_\al \ee where
$N_\al(t)=\int_{0}^{L_\al}{\rho_\al(x,t)dx}$ is the number
operator which counts the electrons which are added to the MZI
path $\alpha$ region at time t, 
$\rho_\al(x,t)=\psi^\dag_\al(x,t)\psi_\al(x,t)$, and
$C_\al=\frac{e^2L_\al}{hv_g}$ is the electric capacitance of path
$\alpha$, induced by the dispersion in $H_0$ and Pauli principle.
Hence, the added terms in the Hamiltonian do not have additional
free parameters.

The role of the additional term is to locally raise the bottom of the conduction
band (the electro-static potential) in path $\alpha$ according to
its overall local charging. The effect is exactly such that when a
full beam between energies $E_F$  and $E_F+\Delta\mu$ enters the path,
the last fully occupied state remains with same momentum $k_F$,
with a new energy $\bar{\mu}=E_F+\Delta\mu$, which restores BLA validity.
We should note that in addition to this dynamic effect, the new
terms also add electron-hole excitations above the Fermi sea,
which we ignore \cite{Mar05}.

\begin{figure}[h]
\centering
\includegraphics[scale=0.4]{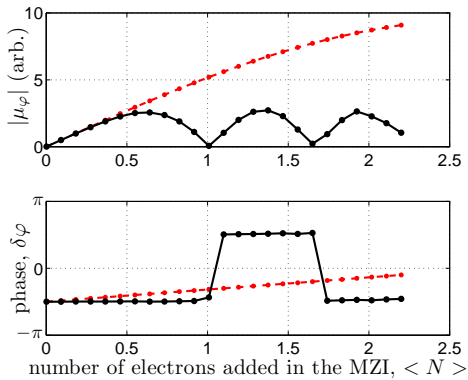}
\caption{The amplitude and phase shift of the interference
oscillations in the chemical potential of the MZI drain, as a
function of $\la N \ra=\overline{L}\Delta\mu/hv_g$ in the non-interacting
model (red curves) and in presence of the interaction (Eq.
(\ref{eq:vis2}), black curves), for $\la \Delta N \ra / \la
N\ra=\Delta L / \overline{L}=0.2$. The lines between the
calculated points are a guide to the eye. Note that we plotted here the total current visibility, which is more suitable for analyzing the lobes then the differential visibility \cite{Rou07}. \label{fig3} }
\end{figure}

\begin{figure}[h]
\centering
\includegraphics[scale=0.4]{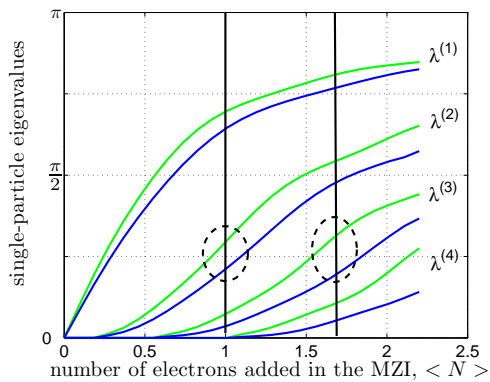}
\caption{The pairs of the single-particle eigenvalues of the phase
operators $w_1$ and $w_2$ of the two paths (blue and green lines), as a function
of $\la N\ra$, for $\la\Delta N\ra/\la N\ra=0.2$. Black
lines mark the zero visibility in Fig. (\ref{fig3}). Note that at
every such zero, there is a pair of eigenvalues whose sum is
$\pi/2$} \label{fig4}
\end{figure}

We shall solve now approximately the equation of motion (EOM) for the
interacting case. Keeping the definition in Eq. \ref{eq:mud2} and
expecting the same functional dependence of $\mu_{D_\al}$
on $\varphi_0$ as in the non-interacting case, we want to show
that $\mu_{\varphi,\al}$
 develops a lobe structure behavior and phase rigidity with an energy scale independent of
$\Delta L$. The EOM for $\psi_\al(x,t)$ reads \bea
\label{eom psi} && \frac{\p\psi_\al(x,t)}{\p
t}=\frac{1}{i\hbar}[\psi_\al(x),H] =
\\ \nonumber && v_g\frac{\p \psi_\al}{\p x}-i\frac{\pi v_g}{L_\al}
\left(\psi_\al(x,t)N_\al+N_\al\psi_\al(x,t)\right)\Theta(x)\Theta(L_\al-x)
\eea Where $\Theta(x)$ is the Heaviside function. We use the
linear relations in  Eq. \ref{cks} to define the operators
$\psi_\al(x,t)$ at $x<0$ and at $x>L_\al$
 \bea \label{boundry} &&
\psi_{(D_\alpha)}(x,t)=\sum_{\beta=1,2}
{s^{(QPC2)}_{\alpha\beta}\psi_{(\beta)}(x,t)}, ~~ x>L_\al
\\ \nonumber &&
\psi_{(\alpha)}(x,t)=\sum_{\beta=1,2}
{s^{(QPC1)}_{\alpha\beta}\psi_{(S_\beta)}(x,t)}, ~~ x<0 \eea where
 $\psi_{(S_\alpha)}(x,t)=\sum_{k}e^{ikx}c_k^{(S_\alpha)}(t)$.
 Having the non-interacting solution
$\psi_\al(x,t)=\psi_\al^{(0)}(x-v_g t)$ at $x<0$ as
boundary condition, Eq. \ref{eom psi} has a unique formal
solution, \be\label{eq:formal sol} \psi_\al(x,t)=\left\{\ba{cl}
\psi_\al^{(0)}(x-v_g t) &x<0\\
\mathcal{T}[U_\al(t)]\psi_\al^{(0)}(x-v_g t)\mathcal{T}^{-1} [U_\al(t)]&    0<x<L_\al\\
\psi_\al(L_\al,t-(x-L_\al)/v_g) & L_\al<x \ea\right. \ee where
$\mathcal{T}$ is time-ordering operator, and we define \be
\label{U-op-1} U_\al(t)= e^{-\frac{i\pi
v_g}{L_\al}\int_{t-x/v_g}^{t}dt' N_\al(t')} \ee Eq. \ref{eq:formal
sol} is difficult to evaluate analytically, due to the fact that
$U_\al(t)$ depends on $\psi_\al(x,t)$. However, one can obtain
some general properties of the highly correlated quantum state at
the drains, using the following derivation. First we approximate
$U_\al(t)$  by replacing $\rho_\al(x,t)$ in $N_\al(t)$  with the
non-interacting density $\rho_\al^0(x-v_gt)$. One should note that
this is a very crude approximation as electrons really repel each
other and change each others group velocity considerately; hance
the true solution might be different in its fine details. However,
note that unlike the MFA, this approximation maintains the
non-Gaussian behavior of the shot-noise, as well as the negative
correlation of the shot-noise between the two paths. Next, due to
the fact that $N^0(t)$ are Bosonic, $[N^0_\al(t), N^0_\al(t')]$ is
a C-number, so the T-ordering can be eliminated by repeatedly
using Baker-Hausdorff formula, such that the phases originating
from the $T$ and the $T^{-1}$ ordering cancel each other. Then, we
take $t=0$ and rearrange the operator in the exponent $
U_\al=U_\al(t=0)= e^{-i\phi_\al}$ as a weighted integral over the
density operator
$\phi_\al=\int_{-\infty}^{\infty}w_\al(x)\rho^0_\al(x)dx$.
Straightforward calculation shows  $w_\al(x)$ to be a triangle \be
\label{triangle} w_\al(x)=\left\{\ba{ll}
  \pi(1-|\frac{x-L_\al}{L_\al}|) & |x-L_\al|<L_\al \\
  0 & |x-L_\al|>L_\al
\ea \right. \ee An additional approximation is to restrict the
energy excitations in the operators $\rho_\al^0(x)$ and therefore
also in the operator $\phi_\al$
 \be \phi_\al \approx
\widetilde{\phi}_\al=\sum_{k,k'=k_F}^{k_F+\Delta\mu/\hbar
v_g}w_{\al,k-k'}c^\dag_{\al,k} c_{\al,k'}\ee where $w_{\al,k-k'}$
is the Fourier transform of $w_\al(x)$, and $\widetilde{O}$ denote
an operator restricted to the energies $[E_F, E_F+\Delta\mu]$. This
approximation is very intuitive as it takes into account only
dephasing due to excess shot-noise, disregarding virtual
transitions to energies above $E_F+\Delta\mu$. It was proven very useful
in explaining the unique experimental results of dephasing by
shot-noise \cite{Ned07b}. With these approximations, the
pre-factor of the phase dependant part of Eq. \ref{eq:mud2} reads
\be \label{eq:vis1} \mu_\varphi= hv_g\langle\Psi_0|
\widetilde{U}_1^\dag \widetilde{U}_2\widetilde\psi_1^{0\dag}
\left(L_1\right)\widetilde\psi_2^0\left(L_2\right)\widetilde{U}_1^\dag
\widetilde{U}_2|\Psi_0\rangle \ee where
$|\Psi_0\ra=\left(\prod_{k=k_F}^{k_F+\Delta\mu/\hbar
v_g}c_k^{(S_1)\dag}\right)|gs(\mu_{S_{1,2}}=E_F)\ra$. Eq.
\ref{eq:vis1} can be evaluated by noting that $\widetilde{\phi}_1$
and $\widetilde{\phi}_2$ commute, so one can introduce a basis of
a complete set of eigenstates of the two operators
$\{|\varphi_1,\varphi_2\rangle\} $. Note that the single-particle
matrix $w_{\al,k,k'}$, restricted to the voltage window $[E_F,
E_F+\Delta\mu]$, has a discrete set of non-vanishing eigenvalues,
corresponding to localized electron-states near the finite
influence region of w(x) (Eq. \ref{triangle}). The eigenstate
$|\varphi_\al\rangle$ of the operator $\widetilde{\phi}_\al$ is a
specific choice of occupations of those single-particle states,
the value $\varphi_\al$ being the sum of the eigenvalues of the
occupied states. With two insertions of such a full set, Eq.
\ref{eq:vis1} now reads
 \bea \label{eq:vis2}
 \mu_\varphi&=&hv_g \sum_{\varphi_1,\varphi_2,\varphi'_1,\varphi'_2}\la\Psi_0|\varphi_1,
\varphi_2\ra\la\varphi_1|\widetilde{\psi}_1^{0\dag}\left(L_1,0\right)|\varphi'_1\ra
\\ \nonumber&&\cdot\la\varphi_2|\widetilde{\psi}_2^0
\left(L_2,0\right)|\varphi'_2\ra\la\varphi'_1,\varphi'_2|\Psi_0\ra\cdot
e^{i(\varphi_1-\varphi_2+\varphi'_1-\varphi'_2)} \eea
 This
expression differs from the non-interacting one only by the
exponent at the end of the r.h.s. of the equation.  Eq.
\ref{eq:vis2} can be evaluated numerically, through
diagonalization of the matrices $w_{\al,k,k'}$.
Fig. \ref{fig3}
shows the result for the visibility, as a function of $\la
N\ra=\la N_1+N_2\ra=\frac{\overline{L}}{hv_g}\Delta\mu$, at $\frac{\la
\Delta N\ra}{\la N\ra}=\frac{\Delta L}{\overline{L}}=0.2$, and
$|t_1|^2=|t_2|^2=0.5$. The prediction of the non-interacting model
is also plotted for comparison.
One can clearly see that while in the non-interacting
model the visibility and phase are slowly varying (because the
scale on which they change is proportional to $\Delta L^{-1}$), in
the interacting model a lobe pattern appears in the visibility,
with a stick-slip behavior of the phase. The lobes evolution is
much faster, apparently proportional to $\la N \ra^{-1}$ and hence
to $\overline{L}^{-1}$, in agreement with the experimental
results. Fig. \ref{fig4} shows the evolution of the eigenvalues of
the single-particle matrices $w_1$ and $w_2$ as a function of $\la
N \ra$.
At small bias there are two non-zero eigenvalues, one for each of
the phase operators.  When put in the exponents of Eq.
\ref{eq:vis2}, they exactly cancel the kinetic phase induced from
$H_0$, resulting in phase rigidity. The first zero of the
visibility occurs at $\la N\ra =1$. As seen from Fig.
\ref{fig4}, at this value the sum of the second pair of
single-particle eigenvalues of the two arms is exactly
$\lambda^{(2)}_1+ \lambda^{(2)}_2=\pi/2$. This can be understood,
by assuming that the operators
$\widetilde{\psi}_1^{0\dag}\left(L_1,0\right)$ and
$\widetilde{\psi}_2^0\left(L_2,0\right)$ in Eq. \ref{eq:vis2} act
effectively only on the first single-particle state (the numeric
calculation indeed shows that the contribution of the other electron
transitions in Eq. \ref{eq:vis2} is negligible). However the phase
of these two operators is modulated by
$\varphi^{(1)}-\varphi^{(2)}+\varphi'^{(1)}-\varphi'^{(2)}$, which
contains also the occupation of the second eigenstates. As the
second electron can occupy the lower arm or the upper arm, a zero
visibility happens when the difference in the added phase between
these two options become $\pi$, so they coherently cancel each
other. Quite generally, the visibility goes to zero whenever a sum of a pair of a single-particle eigenvalues of the two paths reaches $\pi/2$.

Compared with the experiment, the theory still has shortcomings.
The first is that the visibility lobes do not have the overall
decaying envelope. Moreover, differentiating the curve in Fig.
(\ref{fig3}) near $\la N\ra=1$, leads to a phase-dependant
differential response $d\mu_{D_\al,\varphi}/d\mu_{S_1}$  which is
twice that of the linear response; a differential visibility of
200\%. These might be a nonphysical result of our approximations,
which will be absent in the exact solution of Eq. \ref{eom psi}.

In conclusion, in this paper we investigated the effect of quantum shot-noise inside electronic interferometers on the interference visibility in the non-linear regime. It was well established that an interaction Hamiltonian such as in
Eq. \ref{eq:NewHam} must always replace the non-interacting one, which quite generally leads to lobe pattern of the visibility with increasing source bias.  It would be desirable to apply the theory developed here to other
mesoscopic devices such as the two-terminal AB interferometers
\cite{Yac96, Wie03}, the MZI working in the fractional quantum
Hall regime, or the two-particle interferometer
\cite{Sam03,Ned07c}.

We wish to thank A. Stern, Y. Levinson and M. Heiblum for long and
very useful discussions.

\bibliography{bibliography}

\begin{thebibliography}{19}
\expandafter\ifx\csname natexlab\endcsname\relax\def\natexlab#1{#1}\fi
\expandafter\ifx\csname bibnamefont\endcsname\relax
  \def\bibnamefont#1{#1}\fi
\expandafter\ifx\csname bibfnamefont\endcsname\relax
  \def\bibfnamefont#1{#1}\fi
\expandafter\ifx\csname citenamefont\endcsname\relax
  \def\citenamefont#1{#1}\fi
\expandafter\ifx\csname url\endcsname\relax
  \def\url#1{\texttt{#1}}\fi
\expandafter\ifx\csname urlprefix\endcsname\relax\def\urlprefix{URL }\fi
\providecommand{\bibinfo}[2]{#2}
\providecommand{\eprint}[2][]{\url{#2}}

\bibitem[{\citenamefont{Timp et~al.}(1987)\citenamefont{Timp, Chang,
  Cunningham, Chang, Mankiewich, Behringer, and Howard}}]{Tim87}
\bibinfo{author}{\bibfnamefont{G.}~\bibnamefont{Timp}},
  \bibinfo{author}{\bibfnamefont{A.~M.} \bibnamefont{Chang}},
  \bibinfo{author}{\bibfnamefont{J.~E.} \bibnamefont{Cunningham}},
  \bibinfo{author}{\bibfnamefont{T.~Y.} \bibnamefont{Chang}},
  \bibinfo{author}{\bibfnamefont{P.}~\bibnamefont{Mankiewich}},
  \bibinfo{author}{\bibfnamefont{R.}~\bibnamefont{Behringer}},
  \bibnamefont{and} \bibinfo{author}{\bibfnamefont{R.~E.}
  \bibnamefont{Howard}}, \bibinfo{journal}{Phys. Rev. Lett.}
  \textbf{\bibinfo{volume}{58}}, \bibinfo{pages}{2814} (\bibinfo{year}{1987}).

\bibitem[{\citenamefont{Yacoby et~al.}(1996)\citenamefont{Yacoby, Schuster, and
  Heiblum}}]{Yac96}
\bibinfo{author}{\bibfnamefont{A.}~\bibnamefont{Yacoby}},
  \bibinfo{author}{\bibfnamefont{R.}~\bibnamefont{Schuster}}, \bibnamefont{and}
  \bibinfo{author}{\bibfnamefont{M.}~\bibnamefont{Heiblum}},
  \bibinfo{journal}{Phys. Rev. B} \textbf{\bibinfo{volume}{53}},
  \bibinfo{pages}{9583} (\bibinfo{year}{1996}).

\bibitem[{\citenamefont{Buks et~al.}(1998)\citenamefont{Buks, Schuster,
  Heiblum, Mahalu, and Umanski}}]{Buk98}
\bibinfo{author}{\bibfnamefont{E.}~\bibnamefont{Buks}},
  \bibinfo{author}{\bibfnamefont{R.}~\bibnamefont{Schuster}},
  \bibinfo{author}{\bibfnamefont{M.}~\bibnamefont{Heiblum}},
  \bibinfo{author}{\bibfnamefont{D.}~\bibnamefont{Mahalu}}, \bibnamefont{and}
  \bibinfo{author}{\bibfnamefont{V.}~\bibnamefont{Umanski}},
  \bibinfo{journal}{Nature} \textbf{\bibinfo{volume}{391}},
  \bibinfo{pages}{871} (\bibinfo{year}{1998}).

\bibitem[{\citenamefont{Ji et~al.}(2003)\citenamefont{Ji, Chung, Shprinzak,
  Heiblum, Mahalu, and Shtrikman}}]{Jiy03}
\bibinfo{author}{\bibfnamefont{Y.}~\bibnamefont{Ji}},
  \bibinfo{author}{\bibfnamefont{Y.}~\bibnamefont{Chung}},
  \bibinfo{author}{\bibfnamefont{D.}~\bibnamefont{Shprinzak}},
  \bibinfo{author}{\bibfnamefont{M.}~\bibnamefont{Heiblum}},
  \bibinfo{author}{\bibfnamefont{D.}~\bibnamefont{Mahalu}}, \bibnamefont{and}
  \bibinfo{author}{\bibfnamefont{H.}~\bibnamefont{Shtrikman}},
  \bibinfo{journal}{Nature} \textbf{\bibinfo{volume}{422}},
  \bibinfo{pages}{415} (\bibinfo{year}{2003}).

\bibitem[{\citenamefont{Samuelsson et~al.}(2003)\citenamefont{Samuelsson,
  Sukhorukov, and B\"uttiker}}]{Sam03}
\bibinfo{author}{\bibfnamefont{P.}~\bibnamefont{Samuelsson}},
  \bibinfo{author}{\bibfnamefont{E.~V.} \bibnamefont{Sukhorukov}},
  \bibnamefont{and}
  \bibinfo{author}{\bibfnamefont{M.}~\bibnamefont{B\"uttiker}},
  \bibinfo{journal}{Phys. Rev. Lett.} \textbf{\bibinfo{volume}{91}},
  \bibinfo{pages}{157002} (\bibinfo{year}{2003}).

\bibitem[{\citenamefont{Neder et~al.}(2007{\natexlab{a}})\citenamefont{Neder,
  Ofek, Chung, Heiblum, Mahalu, and Umansky}}]{Ned07c}
\bibinfo{author}{\bibfnamefont{I.}~\bibnamefont{Neder}},
  \bibinfo{author}{\bibfnamefont{N.}~\bibnamefont{Ofek}},
  \bibinfo{author}{\bibfnamefont{Y.}~\bibnamefont{Chung}},
  \bibinfo{author}{\bibfnamefont{M.}~\bibnamefont{Heiblum}},
  \bibinfo{author}{\bibfnamefont{D.}~\bibnamefont{Mahalu}}, \bibnamefont{and}
  \bibinfo{author}{\bibfnamefont{V.}~\bibnamefont{Umansky}},
  \bibinfo{journal}{Nature} \textbf{\bibinfo{volume}{448}},
  \bibinfo{pages}{333} (\bibinfo{year}{2007}{\natexlab{a}}).

\bibitem[{\citenamefont{Buttiker}(1988)}]{But88}
\bibinfo{author}{\bibfnamefont{M.}~\bibnamefont{Buttiker}},
  \bibinfo{journal}{IBM J. Res. Dev.} \textbf{\bibinfo{volume}{32}},
  \bibinfo{pages}{63} (\bibinfo{year}{1988}).

\bibitem[{\citenamefont{Neder et~al.}(2006)\citenamefont{Neder, Heiblum,
  Levinson, Mahalu, and Umansky}}]{Ned06}
\bibinfo{author}{\bibfnamefont{I.}~\bibnamefont{Neder}},
  \bibinfo{author}{\bibfnamefont{M.}~\bibnamefont{Heiblum}},
  \bibinfo{author}{\bibfnamefont{Y.}~\bibnamefont{Levinson}},
  \bibinfo{author}{\bibfnamefont{D.}~\bibnamefont{Mahalu}}, \bibnamefont{and}
  \bibinfo{author}{\bibfnamefont{V.}~\bibnamefont{Umansky}},
  \bibinfo{journal}{Phys. Rev. Lett.} \textbf{\bibinfo{volume}{96}},
  \bibinfo{pages}{016804} (\bibinfo{year}{2006}).

\bibitem[{\citenamefont{Roulleau et~al.}(2007)\citenamefont{Roulleau, Portier,
  Glattli, Roche, Cavanna, Faini, Gennser, and Mailly}}]{Rou07}
\bibinfo{author}{\bibfnamefont{P.}~\bibnamefont{Roulleau}},
  \bibinfo{author}{\bibfnamefont{F.}~\bibnamefont{Portier}},
  \bibinfo{author}{\bibfnamefont{D.~C.} \bibnamefont{Glattli}},
  \bibinfo{author}{\bibfnamefont{P.}~\bibnamefont{Roche}},
  \bibinfo{author}{\bibfnamefont{A.}~\bibnamefont{Cavanna}},
  \bibinfo{author}{\bibfnamefont{G.}~\bibnamefont{Faini}},
  \bibinfo{author}{\bibfnamefont{U.}~\bibnamefont{Gennser}}, \bibnamefont{and}
  \bibinfo{author}{\bibfnamefont{D.}~\bibnamefont{Mailly}},
  \bibinfo{journal}{Phys. Rev. B} \textbf{\bibinfo{volume}{76}},
  \bibinfo{eid}{161309} (\bibinfo{year}{2007}).

\bibitem[{\citenamefont{van der Wiel~et. al.}(2003)}]{Wie03}
\bibinfo{author}{\bibfnamefont{W.}~\bibnamefont{van der Wiel~et. al.}},
  \bibinfo{journal}{Phys. Rev. B} \textbf{\bibinfo{volume}{67}},
  \bibinfo{pages}{033307} (\bibinfo{year}{2003}).

\bibitem[{\citenamefont{Sukhorukov and Cheianov}(2007)}]{Suk07}
\bibinfo{author}{\bibfnamefont{E.~V.} \bibnamefont{Sukhorukov}}
  \bibnamefont{and} \bibinfo{author}{\bibfnamefont{V.~V.}
  \bibnamefont{Cheianov}}, \bibinfo{journal}{Phys. Rev. Lett.}
  \textbf{\bibinfo{volume}{99}}, \bibinfo{eid}{156801} (\bibinfo{year}{2007}).

\bibitem[{\citenamefont{Chalker et~al.}(2007)\citenamefont{Chalker, Gefen, and
  Veillette}}]{Cha07}
\bibinfo{author}{\bibfnamefont{J.~T.} \bibnamefont{Chalker}},
  \bibinfo{author}{\bibfnamefont{Y.}~\bibnamefont{Gefen}}, \bibnamefont{and}
  \bibinfo{author}{\bibfnamefont{M.~Y.} \bibnamefont{Veillette}},
  \bibinfo{journal}{Phys. Rev. B} \textbf{\bibinfo{volume}{76}},
  \bibinfo{eid}{085320} (\bibinfo{year}{2007}).

\bibitem[{\citenamefont{Levkivskyi and Sukhorukov}(2008)}]{Lev08}
\bibinfo{author}{\bibfnamefont{I.~P.} \bibnamefont{Levkivskyi}}
  \bibnamefont{and} \bibinfo{author}{\bibfnamefont{E.~V.}
  \bibnamefont{Sukhorukov}}, \bibinfo{journal}{arXiv:condmat/0801.2338}
  (\bibinfo{year}{2008}).

\bibitem[{\citenamefont{Christen and B\"{u}ttiker}(1996)}]{Chr96}
\bibinfo{author}{\bibfnamefont{T.}~\bibnamefont{Christen}} \bibnamefont{and}
  \bibinfo{author}{\bibfnamefont{M.}~\bibnamefont{B\"{u}ttiker}},
  \bibinfo{journal}{EPL (Europhysics Letters)} \textbf{\bibinfo{volume}{35}},
  \bibinfo{pages}{523} (\bibinfo{year}{1996}).

\bibitem[{\citenamefont{Friedel}(1952)}]{Fri52}
\bibinfo{author}{\bibfnamefont{J.}~\bibnamefont{Friedel}},
  \bibinfo{journal}{Philos. Mag.} \textbf{\bibinfo{volume}{43}},
  \bibinfo{pages}{153} (\bibinfo{year}{1952}).

\bibitem[{\citenamefont{Kitagawa and Ueda}(1991)}]{Kit91}
\bibinfo{author}{\bibfnamefont{M.}~\bibnamefont{Kitagawa}} \bibnamefont{and}
  \bibinfo{author}{\bibfnamefont{M.}~\bibnamefont{Ueda}},
  \bibinfo{journal}{Phys. Rev. Lett.} \textbf{\bibinfo{volume}{67}},
  \bibinfo{pages}{1852} (\bibinfo{year}{1991}).

\bibitem[{\citenamefont{Neder et~al.}(2007{\natexlab{b}})\citenamefont{Neder,
  Marquardt, Heiblum, Mahalu, and Umansky}}]{Ned07b}
\bibinfo{author}{\bibfnamefont{I.}~\bibnamefont{Neder}},
  \bibinfo{author}{\bibfnamefont{F.}~\bibnamefont{Marquardt}},
  \bibinfo{author}{\bibfnamefont{M.}~\bibnamefont{Heiblum}},
  \bibinfo{author}{\bibfnamefont{D.}~\bibnamefont{Mahalu}}, \bibnamefont{and}
  \bibinfo{author}{\bibfnamefont{V.}~\bibnamefont{Umansky}},
  \bibinfo{journal}{Nature Physics} \textbf{\bibinfo{volume}{3}},
  \bibinfo{pages}{534} (\bibinfo{year}{2007}{\natexlab{b}}).

\bibitem[{\citenamefont{Neder and Marquardt}(2007)}]{Ned07a}
\bibinfo{author}{\bibfnamefont{I.}~\bibnamefont{Neder}} \bibnamefont{and}
  \bibinfo{author}{\bibfnamefont{F.}~\bibnamefont{Marquardt}},
  \bibinfo{journal}{New. J. Phys.} \textbf{\bibinfo{volume}{9}},
  \bibinfo{pages}{1} (\bibinfo{year}{2007}).

\bibitem[{\citenamefont{Marquardt}(2005)}]{Mar05}
\bibinfo{author}{\bibfnamefont{F.}~\bibnamefont{Marquardt}},
  \bibinfo{journal}{Europhys. Lett.} \textbf{\bibinfo{volume}{72}},
  \bibinfo{pages}{788} (\bibinfo{year}{2005}).

\end{thebibliography}
\end{document}